\documentclass[aps,prx,reprint,superscriptaddress,longbibliography,amssymb,amsmath]{revtex4-1}

\usepackage[%
colorlinks=true,	 
urlcolor=blue,  
linkcolor=blue,  
citecolor=blue
]{hyperref}
\usepackage{color}
\usepackage{xcolor}
\usepackage{graphicx}
\usepackage[normalem]{ulem}
\usepackage{blindtext}

\begin{document}

\title{Coherence and Interaction in confined room-temperature polariton condensates with Frenkel excitons}%

\author{Simon Betzold}
\email{simon.betzold@uni-wuerzburg.de}
\affiliation{Technische Physik, Physikalisches Institut and W\"urzburg‐Dresden Cluster of Excellence ct.qmat, Universit\"at W\"urzburg, D-97074 W\"urzburg, Germany}
\author{Marco Dusel}
\affiliation{Technische Physik, Physikalisches Institut and W\"urzburg‐Dresden Cluster of Excellence ct.qmat, Universit\"at W\"urzburg, D-97074 W\"urzburg, Germany}
\author{Oleksandr Kyriienko}
\affiliation{NORDITA, KTH Royal Institute of Technology and Stockholm University, Roslagstullsbacken 23, SE-106 91 Stockholm, Sweden}
\affiliation{ITMO University, St. Petersburg 197101, Russia}
\author{Christof P. Dietrich}
\affiliation{Technische Physik, Physikalisches Institut and W\"urzburg‐Dresden Cluster of Excellence ct.qmat, Universit\"at W\"urzburg, D-97074 W\"urzburg, Germany}
\author{Sebastian Klembt}
\affiliation{Technische Physik, Physikalisches Institut and W\"urzburg‐Dresden Cluster of Excellence ct.qmat, Universit\"at W\"urzburg, D-97074 W\"urzburg, Germany}
\author{J\"urgen Ohmer}
\affiliation{Department of Biochemistry, Universit\"at W\"urzburg, D-97074 W\"urzburg, Germany}
\author{Utz Fischer}
\affiliation{Department of Biochemistry, Universit\"at W\"urzburg, D-97074 W\"urzburg, Germany}
\author{Ivan A. Shelykh}
\affiliation{Science Institute, University of Iceland, IS-107 Reykjavik, Iceland}
\affiliation{ITMO University, St. Petersburg 197101, Russia}
\author{Christian Schneider}
\email{christian.schneider@uni-wuerzburg.de}
\affiliation{Technische Physik, Physikalisches Institut and W\"urzburg‐Dresden Cluster of Excellence ct.qmat, Universit\"at W\"urzburg, D-97074 W\"urzburg, Germany}
\author{Sven H\"ofling}
\affiliation{Technische Physik, Physikalisches Institut and W\"urzburg‐Dresden Cluster of Excellence ct.qmat, Universit\"at W\"urzburg, D-97074 W\"urzburg, Germany}
\affiliation{SUPA, School of Physics and Astronomy, University of St Andrews, United Kingdom}
\date{Mai 2019}%

\begin{abstract}

The strong light-matter coupling of a microcavity mode to tightly bound Frenkel excitons in organic materials emerged as a versatile, room-temperature compatible platform to study nonlinear many-particle physics and bosonic condensation. However, various aspects of the optical response of Frenkel excitons in this regime remained largely unexplored. Here, we utilize a hemispheric optical cavity filled with the fluorescent protein mCherry to address two important questions in the field of room-temperature polariton condensates. First, combining the high quality factor of the microcavity with a well-defined mode structure allows us to provide a definite answer whether temporal coherence in such systems can become competitive with their low-temperature counterparts. We observe highly monochromatic and coherent light beams emitted from the condensate, characterized by a coherence time greater than 150\,ps, which exceeds the polariton lifetime by two orders of magnitude. Second, the high quality of our device allows to sensibly trace the emission energy of the condensate, and thus to establish a fundamental picture which quantitatively explains the core nonlinear processes yielding the characteristic density-dependent blueshift. We find that the energy shift of Frenkel exciton-polaritons is largely dominated by the reduction of the Rabi-splitting due to phase space filling effects, which is influenced by the redistribution of polaritons in the system. While our finding of highly coherent condensation at ambient conditions addresses the suitability of organic polaritonics regarding their utilization as highly coherent room temperature polariton lasers, shedding light on the non-linearity is of great benefit towards implementing non-linear devices, optical switches, and lattices based on exciton-polaritons at room temperature.
\end{abstract}

\maketitle

\section{Introduction}
Strong coupling between excitons and photons inside a microcavity leads to the formation of cavity polaritons, hybrid light-matter quasiparticles~\cite{Weisbuch1992}. In the low density limit, typically considered in experiments, exciton-polaritons obey bosonic statistics, and thus can undergo a dynamic condensation above a critical particle density due to stimulated processes into a low energy state~\cite{Imamoglu1996}. In the condensed phase, polaritons emit coherent light without requiring population inversion. This process, which was termed polariton lasing, can exhibit a largely decreased threshold as compared to that of conventional photon lasers~\cite{Deng2003}. While the signatures of long range order and first order spatial coherence are routinely observed in polaritonic condensates~\cite{Kasprzak2006,Balili2007}, a pronounced temporal coherence only became accessible by utilizing noise-free pump-lasers~\cite{Krizhanovskii2006,Love2008}, or engineering of a spatial confinement for the system~\cite{Zhang2014,Adiyatulli2015,Klaas2018,Klaas2018a}. 

In the conventional case of GaAs-based microcavity systems, the nonlinear properties originate from the admixture of Wannier-Mott type excitons with large spatial extension. They can interact with each other via the short- and long-range exchange interaction of the excitons~\cite{Ciuti1998,Tassone1999}, leading to an intrinsic non-linearity. This non-linearity has been established as a powerful tool to dynamically manipulate cavity polaritons and their condensates~\cite{Gao2012,Ballarini2013} at ultra-fast time scales~\cite{Hayat2012}, and in complex architectures~\cite{Berloff2017}. Furthermore, it has recently been established as a tool to generate non-classical polariton states via the quantum blockade~\cite{Delteil2019, Munoz-Matutano2019}. 

\begin{figure*}
\includegraphics[width=\textwidth]{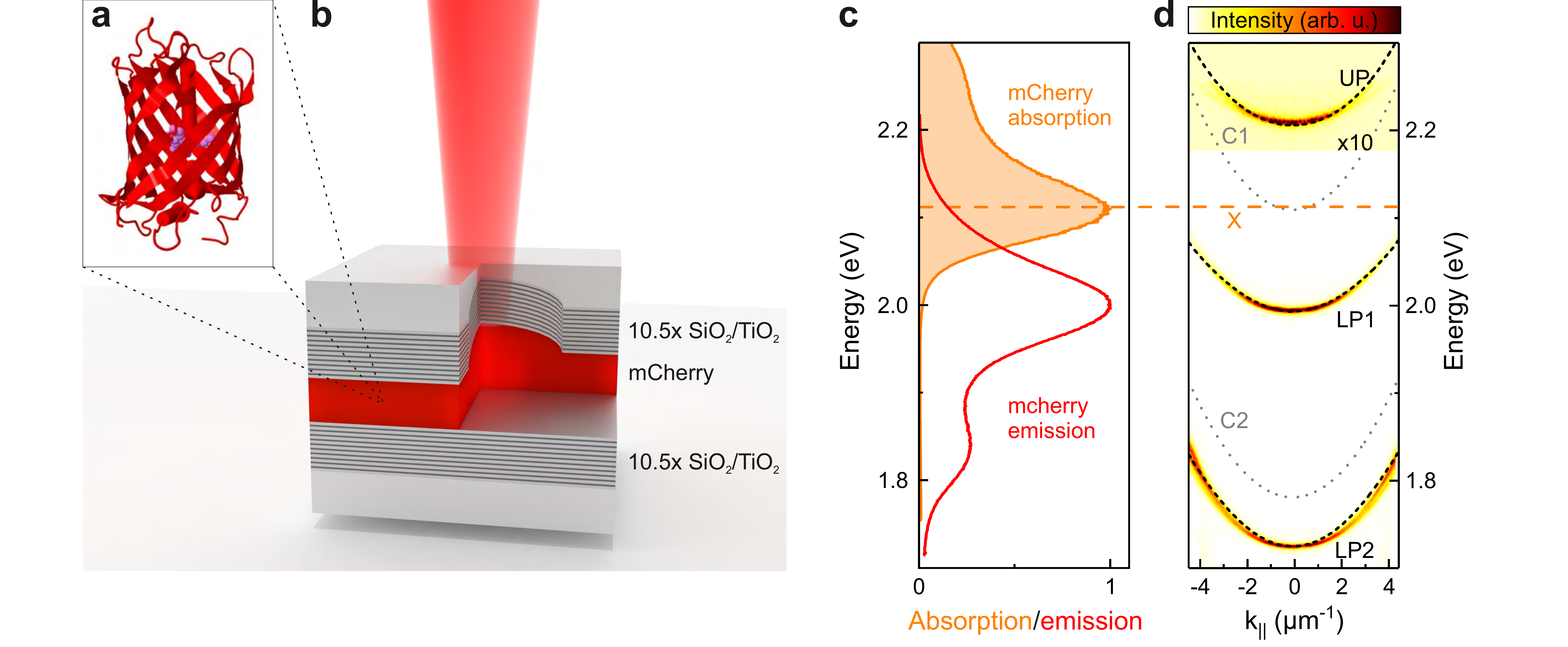}
\caption{\label{fgr:fig1}(a) Schematic image (side view) of the molecular structure of mCherry. The chromophore (violet) is surrounded by an 11-stranded $\beta$-barrel. The image was generated using JMol 14.28.3. (b) Illustration of the microcavity containing a thin film of mCherry. (c) Absorption (orange line) and emission (red line) spectra of a mCherry film with predominant absorption transition at 2.112\,eV (orange dashed line, X). (d) False-color photoluminescence spectrum of a planar area of the device shows two lower polariton branches (LP1, LP2) and an upper polariton branch (UP, magnified in intensity by a factor of 10 for clarity). The position of the polariton branches can nicely be reproduced using a coupled oscillator model (black dashed lines) with a coupling strength $g$ of 133\,meV and two longitudinal photon modes (C1,C2, gray dashed lines) inside the cavity.}
\end{figure*}
Utilization of organic materials in the context of strong light-matter coupling holds enormous potential, as they provide extremely robust polaritons at ambient conditions~\cite{Lidzey1998} and exhibit large coupling strengths up to 1\,eV~\cite{Kena-Cohen2013}. However, signatures of polaritonic condensation and lasing were only reported in a small group of materials hosting tightly bound Frenkel excitons, including melt-grown single crystalline anthracene~\cite{Kena-Cohen2010}, polymers~\cite{Plumhof2013,Daskalakis2014}, fluorescent proteins~\cite{Dietrich2016} and yellow emitting dyes~\cite{Cookson2017}. In a similar manner, polaritonic condensates at ambient conditions were discussed based on wide-bandgap semiconductors, including GaN~\cite{Christopoulos2007,Christmann2008,Bhattacharya2014} and ZnO~\cite{Dai2011,Xu2014} as well as inorganic perovskites~\cite{Su2017}. While in these studies the non-linear behavior in the strong coupling limit was reported, the explanation of the associated coherence of the emitted light and the nature of the interaction processes requires further investigation, both from experimental and theoretical side.

Here, we provide a joint experimental-theoretical study, which allows for a detailed description of the nonlinear response in the organic material-based system. This became possible by utilizing a high quality factor hemispheric cavity filled with the red fluorescent protein mCherry~\cite{Shaner2004}. The unique molecular structure of mCherry proteins allows us to conduct the experiment under quasi-CW condition in the condensation process. Further, the hemispherical cavity provides a sufficiently tight spatial confinement with very high quality factor~\cite{Betzold2017}. This allows us to conduct our experiment in a single mode scenario. Under these conditions, we find strongly extended temporal coherence length exceeding 150\,ps in the condensed phase. We further observe clear indications for both, the contribution of the power-dependent reduction of the Rabi-splitting as well as the redistribution of polaritons in the system, and establish a microscopic model to explain the processes. 

While in most of the reports devoted to polariton condensation with organic materials, optical pumping with short laser pulses (sub-10\,ps) was required to circumvent exciton-exciton annihilation at high excitation densities~\cite{Daskalakis2014}, due to the unique molecular structure fluorescent mCherry can tolerate strongly extended laser pulses. Here, the chromophore is surrounded by a $\beta$-barrel (Fig.~\ref{fgr:fig1}a) which protects the chromophore from the environment. In addition, the barrel-structure effectively reduces concentration-induced exciton-exciton annihilation, even in their solid state and suppresses biomolecular quenching at high excitation densities~\cite{Dietrich2016}.

In conventional semiconductor microcavities the main contribution of the blueshift of a polariton mode is assigned to the Coulomb exchange interaction of Wannier-Mott excitons \cite{Ciuti1998} due to the overlap of the wavefunctions of different excitons. In organic materials such an overlap can not be expected as a result of the strong localition of tightly bound Frenkel excitons. In the following section we provide a theory which accounts for the phase space filling effects for the Frenkel excitons. It predicts a density-dependent reduction of the Rabi-frequency, which consequently results in the blueshift for the lower polariton mode.
\begin{figure*}
\includegraphics[width=\textwidth]{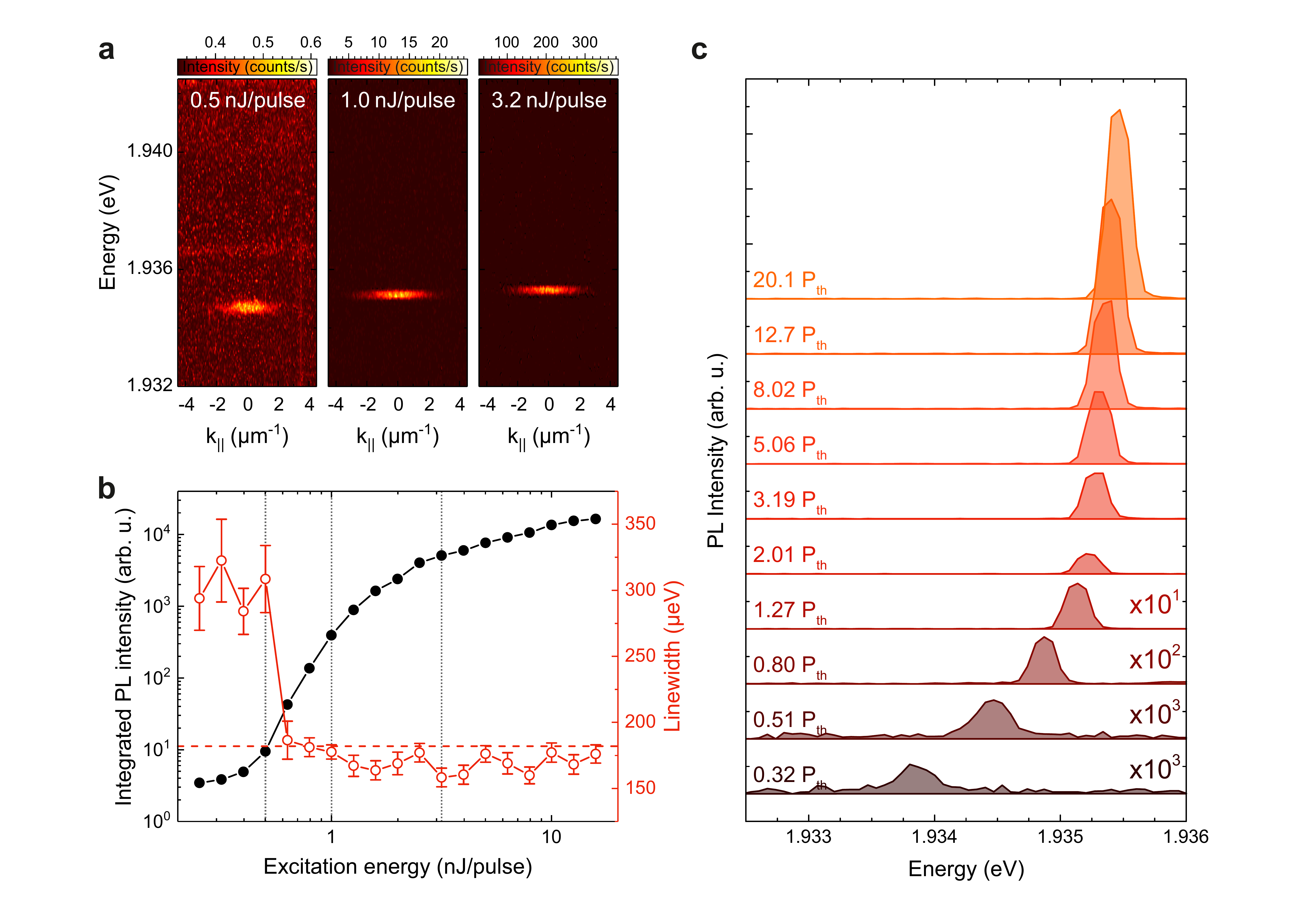}
\caption{\label{fgr:fig2}Excitation power-dependent analysis of the device at the position of a dimple with radius of curvature of $R_C=16\, \mu$m. (a) Far-field resolved spectra for pump powers below the condensation threshold (left, P=0.6\,P$_{th}$), right above the threshold (center, P=1.3\,P$_{th}$) and far past the threshold (right, P=4.1\,P$_{th}$). A significant increase in the intensity and a blueshift of the confined, dispersionless mode is clearly visible. (b) Integrated emission intensity (black filled circles) versus excitation energy. The threshold at P=0.78\,nJ/pulse indicates the onset of polariton condensation of the confined mode. At the same pump power the linewidth collapses to the resolution limit of the spectrometer (180\,$\mu$eV, red dashed line). The gray dotted lines indicate the excitation powers of the spectra shown in (a). (c) Waterfall plot of the PL intensity for different pump powers showing a strongly pronounced blueshift below threshold which continues with a reduced slope after threshold.}
\end{figure*}

\section{Cavity characterization}
A sketch of the studied device is shown in Fig.~\ref{fgr:fig1}b. The system consists of a thin (few hundred nanometers) film of mCherry embedded in a microcavity formed by two distributed Bragg reflectors (DBRs) each consisting of 10.5 alternating pairs of SiO$_2$- and TiO$_2$-layers and with reflectivity $\geq$99.9$\%$ between 1.90\,eV (653\,nm) and 2.13\,eV (582\,nm). For this configuration the mode quality factor exceeds $Q = 2.5\times 10^4$ theoretically and $Q=7.5\times 10^3$ experimentally. Prior to coating the top DBR, we prepared a plateau-like area on the glass substrate with depth and diameter of about 500\,$\mu$m and 4000\,$\mu$m, respectively, by chemical wet etching. On that we have defined lens-shaped indentations using ion beam milling~\cite{Dolan2010,Trichet2015}. These hemi-ellipsoidical dimples have diameters ranging from 2\,$\mu$m to 12\,$\mu$m and depths between 100\,nm and 650\,nm, and the shape of the micro-lenses yields to effective radii of curvature $R_C$ that ranges between 2\,$\mu$m and 360\,$\mu$m. The layer of mCherry is then sandwiched between the two DBRs.

Fig.~\ref{fgr:fig1}c shows the absorption (orange line) and emission (red line) spectra of mCherry. The absorption of mCherry exhibits a predominant transition at 2.112\,eV (587\,nm, orange dashed line, X), whereas the emission has its maximum at 2.003\,eV (619\,nm) with a second smaller maximum at 1.829\,eV (678\,nm).

In Fig. \ref{fgr:fig1}d we plot a false-color photoluminescence (PL) spectrum from the planar area next to the array of lenses. To highlight the behavior of the two-dimensional upper polariton mode, the intensity of the upper area is magnified by a factor of 10 compared to the lower one. For a zero in-plane wave vector $k_{||}=0$ the polariton mode peaks are located at $E_{UP}=2.209$\,eV, $E_{LP1}=1.994$\,eV and $E_{LP2}=1.724$\,eV with linewidths of $w_{UP}=5.9$\,meV, $w_{LP1}=9.3$\,meV and $w_{LP2}=7.4$\,meV. By iteratively calculating the uncoupled photonic modes inside the cavity using transfer-matrix method (gray dotted lines, C1 and C2 are the two longitudinal modes in the cavity) and fitting with a coupled oscillator model (black dashed lines, LP1, LP2 and UP) the positions and the curvatures of the measured polariton dispersions can nicely be reproduced. The cavity parameters can be estimated from the fitting model, which gives an optical cavity length of 1.14\,$\mu$m and a coupling strength $g$ of 133\,meV (Rabi frequency of $\hbar \Omega_{\mathrm{R}} = 2g = 266$\,meV).

\begin{figure*}
\includegraphics[width=\textwidth]{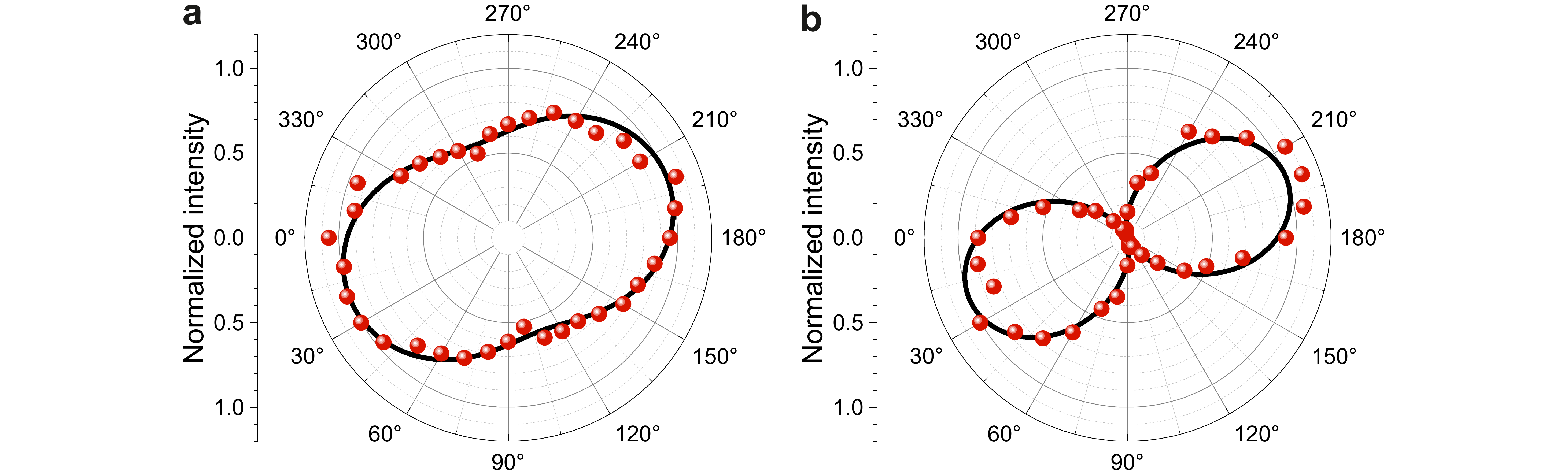}
\caption{\label{fgr:fig3}Polarization-dependent normalized intensities of the hemispherical microcavity below condensation threshold (a) and above (b). The black lines are fits to determine the degree of linear polarization of 0.27(1) below and 0.95(3) above threshold, respectively.}
\end{figure*}

\section{Spectral properties of the polariton condensate}
Next, we provide a detailed power-dependent study of  confined polaritonic modes, associated with the lower polariton branch LP1. We utilized a tunable pulsed laser at a wavelength of 532\,nm (tuned to the first Bragg minimum of the DBR structure), providing pulses with a length of 7\,ns (repetition rate of 2\,Hz), focused to the dimple area with a radius of curvature of $R_C=16\,\mu$m. This extended pulse length provides quasi steady-state conditions, since the lifetime of excitons in mCherry was previously found to be around 1\,ns, and the polaritonic lifetime ranges is around 1\,ps. In Fig.~\ref{fgr:fig2}a we show far-field resolved spectra recorded at different pump powers. Due to the three-dimensional confinement the emission appears dispersion-less. Compared to the spectrum in Fig.~\ref{fgr:fig1}d the mode is slightly shifted to lower energies. This redshift is caused by the larger cavity length due to the dimple depth, which is partly compensated by the spectral blueshift provided by the optical confinement. The calculated optical cavity length in our dimple cavity is 1.21\,$\mu$m allows us to extract polaritonic fractions (so-called Hopfield coefficients), which correspond to an admixture of 23.9\% exciton and 76.1\% photon in the confined LP1 mode. By changing the pump power from 0.50\,nJ (left) over 1.00\,nJ (center) to 3.16\,nJ (right), we can witness a significant increase in the emission intensity and a progressive energy shift of our confined mode. 

The hallmarks of a zero-dimensional polariton laser are illustrated in Figs.~\ref{fgr:fig2},\ref{fgr:fig3}. The threshold behavior of our device is best reflected via the input-output representation depicted in Fig.~\ref{fgr:fig2}b. Here, we find a significant non-linear increase of the emitted intensity at a pump power of 0.78\,nJ/pulse, which is accompanied by a collapse of the linewidth from 300\,$\mu$eV approaching the resolution limit of our optical spectrometer (180\,$\mu$eV). 

Additional proof of the polariton condensation in our system is provided by polarization measurements. Previously, it was found that condensates in planar microcavities \cite{Kasprzak2006, Klopotowski2006,Kasprzak2007} and confined structures ~\cite{Kulakovskii2012,Klaas2019} spontaneously acquire a linear polarization, which is usually pined to some axis defining the anisotropy of the system. In our structure, below threshold, we find a degree of linear polarization of 0.27(1) (see~Fig.~\ref{fgr:fig3}a), which indicates a slight anisotropy of our hemispheric microcavity. Once the threshold is crossed, our condensate acquires a strong degree of linear polarization in excess of 0.95(3), which indicates the efficient relaxation of our system in the lowest, symmetry-split mode. While the strong increase of linear polarization in a weakly split polaritonic resonance is a strong indicator of highly efficient bosonic final state stimulation, it further indicates the possibility to engineer the polarization properties in our hemispheric cavities. This will become of particular interest in coupled cavity structures, where the control of the polariton pseudospin has been identified as a viable quantity in ultra-fast on chip simulation approaches \cite{Dreismann2016}.

Optical nonlinearities in our system become extremely pronounced after reaching the threshold, as it can be seen from the energy-trace of the position of the emission maximum well-captured in the waterfall plot in Fig.~\ref{fgr:fig2}c. We observe that the mode experiences a strongly pronounced blueshift below the condensation threshold, which smoothly changes its slope at threshold and progressively continues up to several tens of the threshold power. This behavior was systematically captured by investigating various hemispheric cavities providing different mode confinements. To quantify the different mode volumes, which result from the different geometries, we performed finite-difference time-domain (FDTD) calculations using the commercial software "Lumerical FDTD solutions" and the mode volume $\displaystyle V_{EM}=\frac{\int{d\mathbf{r} \epsilon \mathbf{E}(\mathbf{r})^2}}{\text{max}(\epsilon \mathbf{E}(\mathbf{r})^2)}$ with $\epsilon$ being a dielectric permittivity and $\mathbf{E}(\mathbf{r})$ being the strength of the electrical field at point $\mathbf{r}$. Empirically, we find that while the general trend of the blueshift scaling with input power is universal (steep slope before the threshold $P_{\mathrm{th}}$, followed by reduced slope after $P_{\mathrm{th}}$), the magnitude of the shift strongly depends on the mode volume. 

In conventional semiconductor microcavities several mechanisms can lead to the blueshift of a polariton mode \cite{Brichkin2011}. It is generally believed that the main impact is provided by the Coulomb exchange interaction of Wannier-Mott excitons \cite{Ciuti1998}, which can be efficient only if there is a substantial overlap of the wavefunctions of different excitons. In organic materials such overlap can not be expected as in this case the tightly bound Frenkel excitons are strongly localized. To explain the origin of the blueshift in the organic cavity under study we provide a theory which accounts for the phase space filling effects for the Frenkel excitons. It predicts the density-dependent decrease of the Rabi-frequency, and consequently leads to the blueshift for the lower polariton mode.

\section{Theoretical description of the blueshift in organic semiconductors}

To describe microscopically the light-matter coupling in the system, we start with the description of a Frenkel exciton as a tightly bound electron-hole pair localized at an atomic site~\cite{Combescot2008}. The electrons and holes are described by $\hat{a}_n$ and $\hat{b}_n$ fermionic annihilation operators, respectively, and obey anticommutation relations $\{ \hat{a}_{n'}, \hat{a}_n^\dagger \} = \{ \hat{b}_{n'}, \hat{b}_n^\dagger \} = \delta_{n',n}$ and $\{ \hat{a}_{n'}, \hat{a}_n \} = \{ \hat{b}_{n'}, \hat{b}_n\} = \{ \hat{a}_{n'}, \hat{b}_n \} = 0$ for any $n,n'$. Here, the operator $\hat{a}_n^\dagger$ creates an electron in the excited state of an atom, and a hole operator $\hat{b}_n^\dagger$ destroys the electron in the ground state (i.e. creates a hole). The site index $n$ runs from one to $N_s$, with the latter being the total number of available sites. The operator corresponding to the creation of the Frenkel exciton $n$ then reads $\hat{X}_n^\dagger = \hat{a}_n^\dagger \hat{b}_n^\dagger$, and acting on vacuum it generates a bound pair at position $n$, $|n\rangle = \hat{a}_n^\dagger \hat{b}_n^\dagger |\text{\o} \rangle = \hat{X}_n^\dagger |\text{\o} \rangle$. The binding energy of the Frenkel exciton comes from electron-hole onsite Coulomb interaction, and is typically large ($>100$\,meV). The intersite Coulomb processes lead to delocalization of a pair, and causes exciton-exciton interaction. Note that contrary to Wannier type excitons the conventional Coulomb exchange interaction between different sites is suppressed due to nearly zero overlap, as the system corresponds to the tight-binding limit. However, the intersite direct terms remain.

The system Hamiltonian can be written as a sum of free energies for the cavity mode $\hat{\mathcal{H}}_{\mathrm{cav}} $ and Frenkel excitons $\hat{\mathcal{H}}_X$, which are strongly coupled due to light-matter interaction. It reads
\begin{align}
\label{eq:H_X+cav}
&\hat{\mathcal{H}} = \hat{\mathcal{H}}_{\mathrm{cav}} + \hat{\mathcal{H}}_X + \hat{\mathcal{H}}_{\mathrm{coupl}} =\\ \notag &= \sum_{\mathbf{k}} \hbar \omega_{c,\mathbf{k}} \hat{c}_{\mathbf{k}}^\dagger \hat{c}_{\mathbf{k}} + \sum_n \Delta_{n} \hat{X}_n^\dagger \hat{X}_n + \sum_{n} G_{n,\mathbf{k}} (\hat{X}_n^\dagger \hat{c}_{\mathbf{k}} + h.c.),
\end{align}
where the cavity mode has the dispersion $\omega_{c,\mathbf{k}}$, with $\hat{c}_{\mathbf{k}}$ ($\hat{c}_{\mathbf{k}}^\dagger$) being bosonic annihilation (creation) operators for a photon at momentum $\mathbf{k}$. The free energy of excitons $\hat{\mathcal{H}}_X$ [second term in Eq.~\eqref{eq:H_X+cav}, second line] is described as a sum of two-level systems with transition energies $\Delta_n$. Finally, the third term $\hat{\mathcal{H}}_{\mathrm{coupl}}$ describing the light-matter coupling provides hybridization for the modes, where $G_{n,\mathbf{k}}$ is a coupling constant which in general can depend on the location of an atom and wavevector of the cavity field.

To describe the coupling of the cavity photon to Frenkel excitons in the momentum space, the latter can be written as a delocalized mode using Fourier transform. The excitonic state reads
\begin{equation}
\label{eq:Fourier}
|X_\mathbf{k}\rangle = \hat{X}_{\mathbf{k}}^\dagger |\text{\o} \rangle = \frac{1}{\sqrt{N_s}} \sum_{n=1}^{N_s} \exp(i \mathbf{k} \cdot \mathbf{r}_n) |n \rangle,
\end{equation}
where $\mathbf{r}_n$ denotes positions of the localized excitations, and $\mathbf{k}$ is an exciton momentum. Using the momentum space description, $\hat{\mathcal{H}}_{\mathrm{coupl}}$ can be rewritten as
\begin{equation}
\label{eq:H_coupl}
\hat{\mathcal{H}}_{\mathrm{coupl}} = \sum_{\mathbf{k}} g_{\mathbf{k}} (\hat{X}_{\mathbf{k}}^\dagger \hat{c}_\mathbf{k} + h.c.),
\end{equation}
where we have redefined the coupling constant $g_\mathbf{k}$ to be dependent on the photon momentum, where the cavity photon is converted into an exciton with the same wavevector.

To describe the behavior of the system, it is instructive to write the Heisenberg equation of motion for the operators. Taking the Frenkel exciton mode with momentum $\mathbf{k}'$, we obtain
\begin{equation}
\label{eq:X_Heisenberg}
i \hbar \frac{\partial \hat{X}_{\mathbf{k}'}}{\partial t} = \left[ \hat{X}_{\mathbf{k}'}, \hat{\mathcal{H}}_{\mathrm{coupl}} \right] = \sum_{\mathbf{k}} g_{\mathbf{k}} \hat{c}_{\mathbf{k}} \left[ \hat{X}_{\mathbf{k}'}, \hat{X}_{\mathbf{k}}^\dagger \right].
\end{equation}
The central object in Eq.~\eqref{eq:X_Heisenberg} is a commutator for a Frenkel exciton mode. For excitons treated as ideal bosons, the commutator equals to the delta function, and only $g_{\mathbf{k}'}$ remains in the $\mathbf{k}$-sum. This corresponds to the linear light-matter coupling term (Rabi-splitting), which is usually considered in the low excitation density limit. We can set its value as a coupling constant taken at the small photon momentum $g_0 \equiv g$. In the case of increased carrier populations the composite nature of excitons, being formed by two fermions, modifies their behavior from being purely bosonic~\cite{Combescot2008a}. Given the specific structure of the Frenkel exciton, its statistics was separately studied~\cite{Combescot2009}. For the localized exciton it can be derived straightforwardly from the electron-hole definition for $\hat{X}_n$, and reads
\begin{equation}
\label{eq:comm_Xn}
\left[ \hat{X}_{n'}, \hat{X}_n^\dagger \right] = \delta_{n',n} - \hat{D}_{n',n} \equiv \delta_{n',n} - \delta_{n',n} (\hat{a}_n^\dagger \hat{a}_n + \hat{b}_n^\dagger \hat{b}_n).
\end{equation}
The last term in Eq.~\eqref{eq:comm_Xn} corresponds to the deviation from commutation relations, and physically denotes the impossibility of double excitation at the single site. Importantly, the presence of density-dependent term $\hat{D}_{n',n}$ makes the light-matter coupling effectively nonlinear.

To find the behavior of the Rabi coupling as a function of the exciton number (related to the pump intensity), we calculate the average of Eq.~\eqref{eq:X_Heisenberg} for the state which contains certain number of Frenkel excitons. This state can be represented by a general density matrix $\rho^X$, which accounts for both thermal and coherent components. The average reads
\begin{eqnarray}
\label{eq:X_rho}
i \hbar \langle \dot{\hat{X}}_{\mathbf{k}'} \rangle =&& \mathrm{Tr} \{ \rho^X \sum_{\mathbf{k}} g_{\mathbf{k}} \langle \hat{c}_{\mathbf{k}} \rangle \left[ \hat{X}_{\mathbf{k}'}, \hat{X}_{\mathbf{k}}^\dagger \right] \}= \\* \nonumber =&& \sum_{\mathbf{k}} g_{\mathbf{k}} \langle \hat{c}_{\mathbf{k}} \rangle \sum_{lm} \rho^X_{lm} \langle N_{m}| \left[ \hat{X}_{\mathbf{k}'}, \hat{X}_{\mathbf{k}}^\dagger \right] | N_l \rangle,
\end{eqnarray}
where we expanded the density matrix in the basis of many-body excitonic states $|N_l\rangle$ with index $l$ denoting certain distribution over the momentum index. Here, $\rho^X_{lm}$ corresponds to the occupations and coherences of the density matrix. The many-body states explicitly read $|N_l\rangle = \hat{X}_{\mathbf{k}_1}^{\dagger n_1^l} \hat{X}_{\mathbf{k}_2}^{\dagger n_2^l} .. \hat{X}_{\mathbf{k}_S}^{\dagger n_S^l} |\textbf{\o}\rangle / \mathcal{N}$. $\mathcal{N}^2 := \langle N_l | N_l \rangle$ corresponds to the normalization constant, and $n_j^l$ are numbers of excitons in the momentum mode $\mathbf{k}_j$.
Exploiting the modified commutation relations for Frenkel excitons in the momentum space~\cite{Combescot2009} and considering diagonal part of the density matrix, Eq.~\eqref{eq:X_rho} yields (see Appendix for the full derivation)
\begin{align}
\label{eq:Rabi_fin}
&i \hbar \langle \dot{\hat{X}}_{\mathbf{k}'} \rangle = \sum_{\mathbf{k}} g_{\mathbf{k}} \langle \hat{c}_{\mathbf{k}} \rangle \sum_{l} \rho^X_{ll} \langle N_{l}| \left[ \hat{X}_{\mathbf{k}'}, \hat{X}_{\mathbf{k}}^\dagger \right] | N_l \rangle \approx \\ \nonumber &\approx g_{\mathbf{k}'} \langle \hat{c}_{\mathbf{k}'} \rangle - 2 g_{\mathbf{k}'} \langle \hat{c}_{\mathbf{k}'} \rangle \sum_l \rho^X_{ll} \frac{(n_1^l + n_2^l + .. + n_S^l)}{N_s} = \\ \nonumber &= g_{\mathbf{k}'} \langle \hat{c}_{\mathbf{k}'} \rangle \Big( 1 - \frac{2 N^X_{\mathrm{tot}}}{N_s} \Big),
\end{align}
where $N^X_{\mathrm{tot}}$ corresponds to the total number of Frenkel excitons, summed over distribution in the momentum space. In Eq.~\eqref{eq:Rabi_fin} we accounted for the linear term in $N_s^{-1}$ only, which corresponds to the lowest density-dependent correction to the light-matter coupling. It is important to note that since we assume all excitons to couple to the cavity mode in the Hamiltonian~\eqref{eq:H_coupl}, the exciton number $N^X_{\mathrm{tot}} = N^X_{\mathrm{PL}}$ corresponds only to excitons coupled to light. Eq.~\eqref{eq:Rabi_fin} thus describes the decrease of the light-matter coupling for the increased total population of Frenkel excitons, as long as they lie within the light cone. The exciton density dependence of the light-matter coupling thus introduces the optically nonlinear behavior, which was observed experimentally.
\begin{figure}
\includegraphics[width=0.4\textwidth]{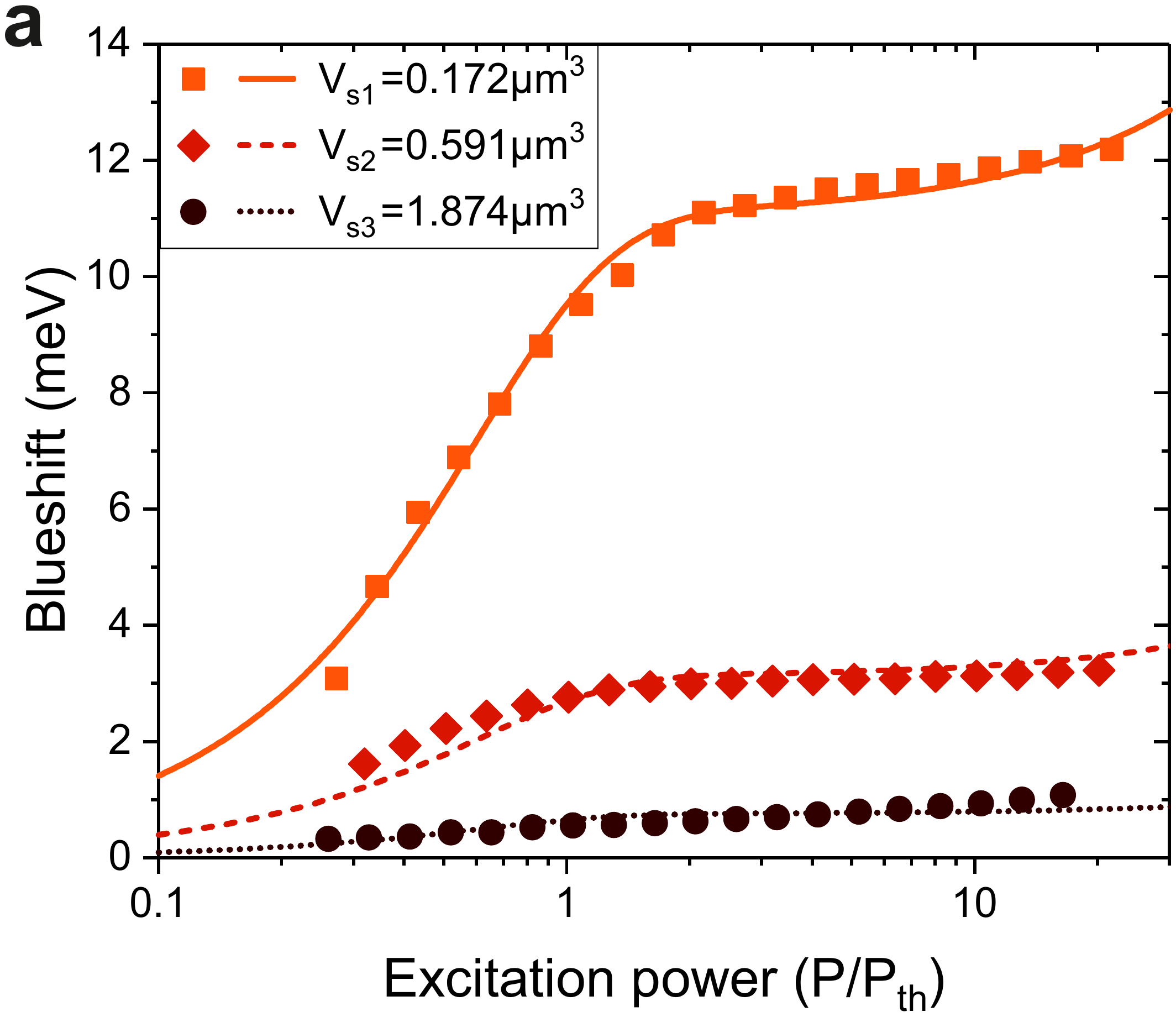}
\caption{\label{fgr:fig4} Excitation power-dependent blueshift for different sample geometries and therefore different mode volumes. The experimental data (symbols) and the model derived in this paper (lines) show very good accordance. The fits were performed using Eq.~\eqref{eq:dE_L} and Eq.~\eqref{eq:N_X_fit} with shared parameters for the 'filling factor' (connected via the mode volume) and the fitting coefficients $\beta_P$ and $c_P$.}
\end{figure}

We can now apply the microscopic theory to the experiment, and compare results. For this, the measured value of a total number of excitons is required. In general, its is a non-trivial function of the pump intensity, as it can depend much on the scattering processes~\cite{Carusotto2013,Byrnes2014}, leading to non-radiative decay or redistribution into modes outside of the light cone. At the same time, we can try to extract this quantity from the PL measurements, where we observe the linear increase before threshold, and the weaker increase after the condensation point. Using Eq.~\eqref{eq:Rabi_fin} and considering small photon momenta $g_{k'=0} \equiv g$, the coupling reads as
\begin{equation}
\label{eq:Omega_R_N}
g (N^X_{\mathrm{PL}}) = g(0) \left[ 1 - \frac{2 N_X^{\mathrm{PL}}}{N_s} \right],
\end{equation}
where by $g(0)$ we explicitly denote the low density coupling term. Taking the lower polariton branch, its blueshift as a function of exciton number is given by
\begin{equation}
\label{eq:dE_L}
\Delta E_{L} (N^X_{\mathrm{PL}}) = \frac{4 g^2(0)}{ \sqrt{\Delta^2 + 4 g^2(0)}} \frac{N^X}{N_s},
\end{equation}
where $\Delta$ is a cavity-exciton detuning. The blueshift value is therefore larger for samples with increased Rabi frequency, while it also has a detuning dependence. It is also inversely proportional to the number of available Frenkel exciton sites, given that the pumping power is fixed.

Note, that the blueshift given by the Eq. \eqref{eq:dE_L} increases linearly with the number of the excitons, while originating from the exciton-photon coupling term. Thus, the effective Hamiltonian describing the process is $\sum_{\mathbf{k},\mathbf{k}'} \hat{c}^\dagger_{\mathbf{k}} \hat{X}_{\mathbf{k}} \hat{X}_{\mathbf{k}'}^\dagger \hat{X}_{\mathbf{k}'} + h.c.$. At the same time, to compare it with the Kerr-type nonlinearity for excitons, typically coming from the Coulomb exchange, we can assign an equivalent Hamiltonian for interacting excitons in bosonized picture characterized by the operators $[\hat{x}_{\mathbf{k}}, \hat{x}_{\mathbf{k}'}^\dagger] = \delta_{\mathbf{k},\mathbf{k}'}$ with interaction Hamiltonian given by
\begin{equation}
\label{eq:H_int}
\hat{H}_{int}=\frac{\hbar u}{2} x^\dagger x^\dagger x x
\end{equation}
where the effective interaction constant reads
\begin{equation}
\label{ueff}
\hbar u = \frac{4 g^2(0)}{ N_s\sqrt{\Delta^2 + 4 g^2(0)}},
\end{equation}
and we considered the exciton to be in the lowest mode $\hat{x}_{0} = : \hat{x}$.

To test the validity of the proposed mechanism, we compare the theory to experiments performed using three mCherry samples put into different optical resonators. The first resonator s2 corresponds to the device already shown in Fig.~\ref{fgr:fig2}, while the other two have either a a hemisphere with smaller diameter (2\,$\mu$m, s1) or a deeper hemisphere (650\,nm, s3). These configurations lead to different radii of curvature and electromagnetic mode volumes. The corresponding electromagnetic mode volumes were estimated by FDTD calculations as $V_{\mathrm{EM}}^{\mathrm{(s1)}} = 0.172\,\mu$m$^3$, $V_{\mathrm{EM}}^{\mathrm{(s2)}} = 0.591\,\mu$m$^3$ and $V_{\mathrm{EM}}^{\mathrm{(s3)}} = 1.874\,\mu$m$^3$.
Let us consider only the dominant light-matter coupling, which occurs between excitons and the C$_1$ mode, and estimate the ratio between blueshifts for samples $\mathrm{S1}$ and $\mathrm{S2}$ using Eq.~\eqref{eq:dE_L}. The Rabi coupling at low excitation power corresponds to 
\begin{equation}
\label{eq:Omega_X_scaling}
g(0) = |\mathbf{d}_{X} \cdot \mathbf{E}| \sqrt{N_s} = |\mathbf{d}_{X} \cdot \mathbf{e}_{\mathrm{EM}}| \sqrt{\frac{\hbar \omega_{c,0}}{2\varepsilon_0 V_{\mathrm{EM}}}} \sqrt{n_s V_s},
\end{equation}
 where $\mathbf{d}_{X}$ denotes the dipole moment for Frenkel excitons (assumed to be pointing same direction) and $\mathbf{E}$ is the electric field for the cavity mode, directed along vector $\mathbf{e}_{\mathrm{EM}}$. Here, $\varepsilon_0$ is the vacuum permittivity, $n_s$ is the density of the molecules available for the coupling and situated in volume $V_s$. 
 From Eq.~\eqref{eq:Omega_X_scaling} one can see that in the crude approximation of the full volume of EM mode being filled with mCherry molecules, $V_{\mathrm{EM}} = V_s$, the Rabi frequency is volume-independent, and has only weak energy dependence for two samples, which can be safely neglected. This then can explain equal Rabi frequencies for samples of different volume.
 
 We consider the same excitation conditions (i.e. number of generated excitons $N_X$), as in the performed experiment. The ratio of blueshifts for different cavities reads
 \begin{equation}
\label{eq:dE_ratio}
\frac{\Delta E_L^{\mathrm{(s2)}}}{\Delta E_L^{\mathrm{(s1)}}} = \left( \frac{g^{\mathrm{(s2)}}}{g^{\mathrm{(s1)}}} \right)^2 \frac{\sqrt{ \Delta_{\mathrm{s1}}^2 + (2g^{\mathrm{(s1)}})^2}}{ \sqrt{ \Delta_{\mathrm{s2}}^2 + (2g^{\mathrm{(s2)}})^2} } \times \frac{V_s^{\mathrm{(s1)}}}{V_s^{\mathrm{(s2)}}}.
 \end{equation}
The analysis of Eq.~\eqref{eq:dE_ratio} gives a prefactor coming from the ratio coupling parameters, times the dominant term corresponding to the ratio of available Frenkel exciton states. For the considered case of equal Rabi frequencies $\Omega_X^{\mathrm{(s2)}} = \Omega_X^{\mathrm{(s1)}} = 266$\,meV, and detunings $\Delta_{\mathrm{s1}} = -97$\,meV, $\Delta_{\mathrm{s2}} = -66$\,meV, the prefactor is close-to-unity. Considering that each mode volume is filled with molecules and couples to light ($V_s^{\mathrm{(s1)}} = V_{\mathrm{EM}}^{\mathrm{(s1)}}$ and $V_s^{\mathrm{(s2)}} = V_{\mathrm{EM}}^{\mathrm{(s2)}}$), the ratio of detunings is $\mathcal{R}_{\mathrm{theory}} = \Delta E_L^{\mathrm{(s2)}}/\Delta E_L^{\mathrm{(s1)}} = 3.55$ and stays the same for equal pump (exciton population).

Finally, let us consider the exciton number dependence as a function of pump power $P$, motivated by the experimental findings. The reasoning comes from the observed PL data where the growth of the intensity before and after the threshold has different rates. This is parametrized by the function
\begin{equation}
\label{eq:N_X_fit}
\frac{N_{X}(P)}{N_{X}(P_{th})} = \frac{\tanh(\beta_P P/P_{th}) + c_P P/P_{th}}{\tanh(\beta_P) + c_P},
\end{equation}
where $\beta_P$ and $c_P$ are dimensionless coefficients. The hyperbolic tangent in the numerator is responsible for the linear growth at small powers. Using the estimated values for the Rabi-splitting, detuning, and mode volumes, the plotted data shown in Fig.~\ref{fgr:fig4} can simultaneously be fitted using Eq.~\eqref{eq:dE_L} and Eq.~\eqref{eq:N_X_fit} and shared parameters for the 'filling factor' (which is connected via the mode volumes for the different samples) and the coefficients $\beta_P$ and $c_P$. In addition, we have used an offset for each data set to account for the blueshift for lower excitation powers. The fits for all three data sets are in very good agreement, even though several approximations were taken. The extracted filling factor at threshold is 1.1\% for the s2 sample and the fitting coefficients are $\beta_P=1.3$ and $c_P=5.5\times 10^{-3}$.

\begin{figure*}
\includegraphics[width=\textwidth]{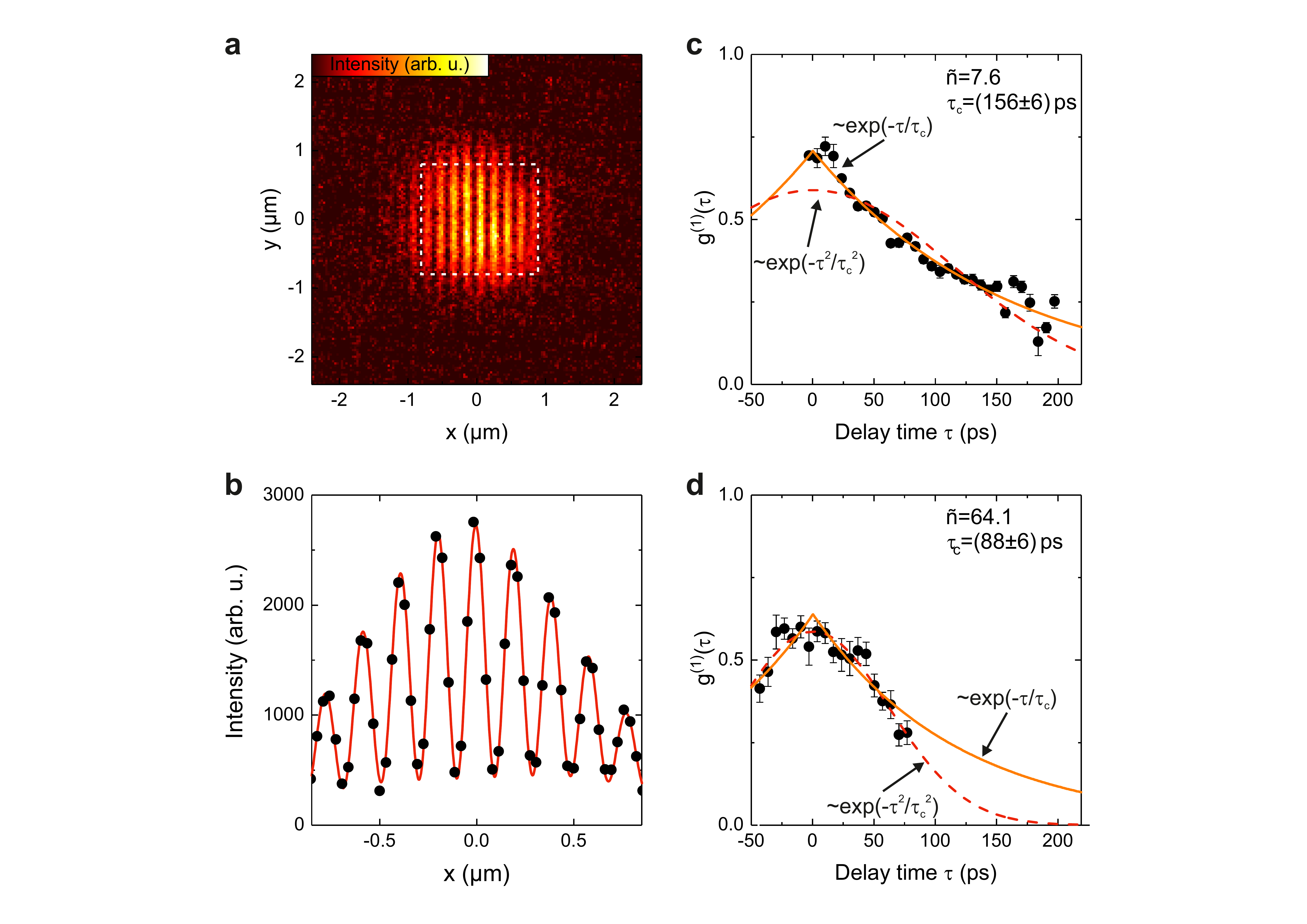}
\caption{\label{fgr:fig5}First-order coherence properties of the polariton condensate. (a) Real-space interference image of the condensate near zero delay time for a pump power of $P=1.7\,P_{th}$. (b) Fringe pattern of the data received by integrating the marked area shown in (a) in y-direction (black dots). The red line is a fit using equation \eqref{eq:visibility} to obtain $g^{(1)} (\tau)$. (c,d) Measured first-order coherence $g^{(1)} (\tau)$ for different delay times (black dots), exponential fits (orange solid lines) and Gaussian fits (red dashed lines) for a pump power of $P=1.7\,P_{th}$ (c) and $P=5.3\,P_{th}$ (d), respectively.  }
\end{figure*}

\section{First-order Coherence properties and interaction strength of the polariton condensate}
The confinement, provided by the hemispheric microcavity, results in a mono-mode condensate, and provides ideal conditions to study the temporal coherence of the emitted light in greater detail. The first-order correlation function $g^{(1)} (\tau)$ describes the temporal phase coherence. From the linewidth a temporal coherence length above threshold of at least 7\,ps can be estimated, yet, with a clear limitation by the instrument response for all data points above threshold. To analyze the coherence build up more in detail we measured the temporal coherence using a Michelson interferometer in mirror-retro-reflector configuration. Since the mirror is mounted on a piezo stage as well as on a motorized linear stage, we can change the delay time $\tau$ between the two arms of the interferometer. Exemplary interference fringes are shown in Fig.~\ref{fgr:fig5}a and Fig.~\ref{fgr:fig5}b, which signify coherence throughout the mode in the dimple cavity. Importantly, by varying the delay time in our interferometer, we can analyze the decay of the fringe amplitude, and fitting the fringe patterns we deduce $g^{(1)}(\tau)$ over $\tau$~\cite{Kim2016} and therewith the coherence time $\tau_{c}$. 

Below threshold a coherence time of $\tau_{c}=(1.5\pm 0.1)$\,ps was determined, which already shows the capability of the device due to the high quality three-dimensional confinement. Above threshold we observe different behavior of  $g^{(1)} (\tau)$ for different occupancies of the condensate. At the pump power of $P=1.7\,P_{th}$, which corresponds to a mean occupancy of $\tilde{n}$=7.6 polaritons, an exponential decay of $g^{(1)} (\tau)\sim \text{exp}(-\tau/\tau_c)$ and a coherence time of $\tau_c$=(156$\pm$6)\,ps was found (see Fig. \ref{fgr:fig5}c). The single exponential decay indicates single-mode condensation as well as predominant intrinsic dephasing of the condensate\cite{Kim2016}. We note, that the coherence time of 156\,ps is unprecedented in any room-temperature polariton condensate, and reflects the maturity of our device. The values reported so far have been limited to a few ps in planar ($\approx 2$\,ps)~\cite{Plumhof2013} and confined ($\approx 1$\,ps)~\cite{Scafirimuto2017} organic samples as well as in planar GaN ($\approx 0.7$\,ps or $\approx 2.8$\,ps electrically driven)~\cite{Christopoulos2007,Bhattacharya2014} and ZnO nanowires ($\approx 1.2$\,ps)~\cite{Xu2014}.

For an increased number of the polaritons in the condensate $\tilde{n}$=64 ($P=5.3\,P_{th}$) the shape of the decay changes from exponential to Gaussian decay $g^{(1)} (\tau)\sim \text{exp}(-\tau^2/\tau_c^2)$ and the coherence length drops to $\tau_c$=(88$\pm$6)\,ps (see Fig. \ref{fgr:fig5}d). The Gaussian broadening has previously been associated with interaction induced energy variations due to number fluctuations in the condensate~\cite{Whittaker2009}, which can be described by the interaction constant $u$ via exp($-\tau^2/\tau_c^2$)=exp(-2$\tilde{n} u^2 \tau^2$). In our case, as it was already discussed, Coulomb-based exciton-exciton interactions are absent and nonlinearity originates from the quenching of the Rabi splitting. However, according to Eq.~\eqref{eq:dE_L} in bosonized picture the latter can be reduced to the effective polariton-polariton interaction with characteristic constant given by Eq.~\eqref{ueff}.  From the fit we deduce $u=(1.0\pm 0.3) \cdot 10^{-3}$\,ps$^{-1}$ or $\hbar u A_{c}=(1.4 \pm 0.4)\,\mu$eV$\times\mu$m$^2$ per polariton in the condensate, which is in excellent agreement with the analysis provided by tracing the condensate energy with pump power. Here, the size of the condensate $A_{c}$ was extracted from spatial resolved PL measurement.

\section{Conclusion}

With our work, we addressed two of the most significant questions in room-temperature polaritonics: The question of coherence, as well as the origin of interaction provided by tightly bound Frenkel excitons in the strong coupling regime. This study became feasible by implementing a high quality factor hemispheric, single mode microcavity, filled with mCherry proteins that can be driven under quasi equilibrium conditions. 

Our core finding reflects that it is possible to obtain very long coherence times in excess of 150\,ps in room-temperature polaritonic condensates. Furthermore, while the energy shift of localized Frenkel exciton polaritons is widely dominated by the quasi-two level structure of the excitons, yielding a progressive reduction of the Rabi-splitting above threshold, which acts analogously to a Coulomb exchange interaction with a strength of approx. $(1.4\pm 0.4)\,\mu$eV~$\times$~$\mu$m$^2$/ polariton.

While our finding of highly coherent condensation at ambient conditions addresses the suitability of organic polaritonics regarding their utilization as highly coherent room temperature polariton lasers, shedding light on the non-linearity will be of great benefit towards implementing non-linear devices, optical switches, and lattices based on exciton-polaritons at room temperature.  
	
\section{Methods}
The sample was excited either by the 532\,nm line of a continuous-wave diode laser that is resonant with the first Bragg minimum of the top mirror for pre-characterization or by a wavelength-tunable optical parametric oscillator system with ns-pulses tuned to 532\,nm for the experiments in the non-linear regime. The emission was collected in reflection geometry using a high-NA objective (50x, NA=0.42) in front of the sample, filtered by using a 550\,nm longpass filter and monitored onto the entrance slit of a 500\,mm spectrometer with a spectral resolution of about 180\,$\mu$eV for energies around 2\,eV. Most measurements were performed in Fourier imaging configuration with a Fourier lens collecting the angle-dependent information in the back-focal plane of the microscope.
The absorption spectrum was measured in transmission configuration using a collimated white light source that was focused onto the backside of a thin film of mCherry coated an a quartz slide. The transmitted light was then collected and recorded with the previously described objective and spectrometer, respectively.

The visibility of the interference fringes for two interfering beam profiles $I_{1}(x)$ and $I_{2}(x)$ is given by~\cite{Kim2016} 
\begin{eqnarray}
\label{eq:visibility}
I(x,\tau)&=&I_{1}(x)+I_{2}(x)+\\* \nonumber &+& 2|g^{(1)} (\tau)|\sqrt{I_{1}(x) I_{2}(x)} \cos \left( \frac{2\pi\theta}{\lambda_{0}}x+\phi \right)
\end{eqnarray}
with 
$\lambda_{0}$ the wavelength of the lasing mode, and $\phi$, $\theta$, $\tau$ the phase difference, angle, and time delay between $I_{1}(x)$ and $I_{2}(x)$. 

\section{Acknowledgements} 
OK and IAS acknowledge support from the ministry of education and science of Russian Federation, projects 14.Y26.31.0015 and  3.2614.2017/4.6. The W\"urzburg group acknowledges financial support from the DFG through the W\"urzburg‐Dresden Cluster of Excellence on Complexity and Topology in Quantum Matter ‐‐ \textit{ct.qmat} (EXC 2147, project‐id 39085490).

\section*{Appendix}

To derive Eq.~\eqref{eq:Rabi_fin} in the main text, we reformulated the deviation from bosonic behavior in the momentum space. The commutator generally has the same form $\left[ \hat{X}_{\mathbf{k}'}, \hat{X}_\mathbf{k}^\dagger \right] = \delta_{\mathbf{k}',\mathbf{k}} - \hat{D}_{\mathbf{k}',\mathbf{k}}$, where the deviation operator can be conveniently defined through its action on the creation operators as~\cite{Combescot2009}
\begin{equation}
\label{eq:comm_D}
\left[ \hat{D}_{\mathbf{k}',\mathbf{k}}, \hat{X}_{\mathbf{p}}^\dagger \right] = \frac{2}{N_s} \hat{X}_{\mathbf{p} + \mathbf{k} - \mathbf{k}'}^\dagger.
\end{equation}
The latter can be seen as a consequence of nonzero portion of excitations as compared to the total number of states. Similarly, taking the commutator for creation of $N$ excitons, the commutator can be derived recursively as
\begin{equation}
\label{eq:comm_D^N}
\left[ \hat{D}_{\mathbf{k}',\mathbf{k}}, (\hat{X}_{\mathbf{p}}^\dagger)^N \right] = \frac{2 N}{N_s} \hat{X}_{\mathbf{p} + \mathbf{k} - \mathbf{k}'}^\dagger (\hat{X}_{\mathbf{p}}^\dagger)^{N-1}.
\end{equation}
The equation of motion for exciton operator Eq.~\eqref{eq:X_rho} taken in a state $\rho^X$ can separated into diagonal and off-diagonal parts as
\begin{align}
\label{eq:X_rho_diag}
&i \hbar \langle \dot{\hat{X}}_{\mathbf{k}'} \rangle = \sum_{\mathbf{k}} g_{\mathbf{k}} \langle \hat{c}_{\mathbf{k}} \rangle \sum_{l} \rho^X_{ll} \langle N_{l}| \left[ \hat{X}_{\mathbf{k}'}, \hat{X}_{\mathbf{k}}^\dagger \right] | N_l \rangle+ \\ \nonumber &+ \sum_{\mathbf{k}} g_{\mathbf{k}} \langle \hat{c}_{\mathbf{k}} \rangle \sum_{l\neq m} \rho^X_{lm} \langle N_{m}| \left[ \hat{X}_{\mathbf{k}'}, \hat{X}_{\mathbf{k}}^\dagger \right] | N_l \rangle,
\end{align}
and we consider two terms separately. The diagonal part is responsible for the leading processes which involve occupations, and are present both before and after condensation. The off-diagonal terms are only present after the condensation, where coherence $\rho^X_{lm}$ ($l \neq m$) is developed.

The diagonal part of the commutator reads
\begin{eqnarray}
\label{eq:Rabi_diag}
&&\sum_{\mathbf{k}} g_{\mathbf{k}} \langle \hat{c}_{\mathbf{k}} \rangle \sum_{l} \rho^X_{ll} \langle N_{l}| \left[ \hat{X}_{\mathbf{k}'}, \hat{X}_{\mathbf{k}}^\dagger \right] | N_l \rangle=  \\*&& \nonumber = g_{\mathbf{k}'} \langle \hat{c}_{\mathbf{k}'} \rangle - \sum_{\mathbf{k}} g_{\mathbf{k}} \langle \hat{c}_{\mathbf{k}} \rangle \sum_l \rho^X_{ll} \frac{1}{\langle N_l | N_l \rangle} \times \\* \nonumber && \times \Big( \langle \text{\o} | \hat{X}_{\mathbf{k}_S}^{n_S^l} .. \hat{X}_{\mathbf{k}_2}^{n_2^l} \hat{X}_{\mathbf{k}_1}^{n_1^l} \left[ \hat{D}_{\mathbf{k}',\mathbf{k}}, \hat{X}_{\mathbf{k}_1}^{\dagger n_1^l} \right]   \hat{X}_{\mathbf{k}_2}^{\dagger n_2^l} .. \hat{X}_{\mathbf{k}_S}^{\dagger n_S^l} |\text{\o} \rangle + \\* \nonumber && + \langle \text{\o} | \hat{X}_{\mathbf{k}_S}^{n_S^l} .. \hat{X}_{\mathbf{k}_2}^{n_2^l} \hat{X}_{\mathbf{k}_1}^{n_1^l} \hat{X}_{\mathbf{k}_1}^{\dagger n_1^l} \left[ \hat{D}_{\mathbf{k}',\mathbf{k}}, \hat{X}_{\mathbf{k}_2}^{\dagger n_2^l} \right]  .. \hat{X}_{\mathbf{k}_S}^{\dagger n_S^l} |\text{\o} \rangle + ... \Big),
\end{eqnarray}
where the first term in the RHS $g_{\mathbf{k}'} \langle \hat{c}_{\mathbf{k}'} \rangle$ corresponds to the coupling in the weak excitation limit. It directly results into a linear light-matter coupling term of $g_0 \equiv g$, where we consider only the modes coupled to light, and set $k' = 0$. The second term then corresponds to the deviation of statistics for the excited system. Its parts depend on the commutation of deviation operator with different momentum states (taken to be $\{\mathbf{k}_1,\mathbf{k}_2, ..., \mathbf{k}_S \} $), weighted with a probability distribution. Looking into the first commutator as an example, we see that
\begin{widetext}
\begin{eqnarray}
\label{eq:diag_Dn_comm}
&&\sum_{\mathbf{k}} g_{\mathbf{k}} \langle \hat{c}_{\mathbf{k}} \rangle \sum_l  \rho^X_{ll} \frac{1}{\langle N_l | N_l \rangle} \langle \text{\o} | \hat{X}_{\mathbf{k}_S}^{n_S^l} .. \hat{X}_{\mathbf{k}_2}^{n_2^l} \hat{X}_{\mathbf{k}_1}^{n_1^l} \left[ \hat{D}_{\mathbf{k}',\mathbf{k}}, \hat{X}_{\mathbf{k}_1}^{\dagger n_1^l} \right]   \hat{X}_{\mathbf{k}_2}^{\dagger n_2^l} .. \hat{X}_{\mathbf{k}_S}^{\dagger n_S^l} |\text{\o} \rangle = \\* \nonumber = &&\sum_{\mathbf{k}} g_{\mathbf{k}} \langle \hat{c}_{\mathbf{k}} \rangle \sum_l \rho^X_{ll} \frac{1}{\langle N_l | N_l \rangle} \frac{2 n_1^l}{N_s} \langle \text{\o} | \hat{X}_{\mathbf{k}_S}^{n_S^l} .. \hat{X}_{\mathbf{k}_2}^{n_2^l} \hat{X}_{\mathbf{k}_1}^{n_1^l}  \hat{X}_{\mathbf{k}_1+\mathbf{k}-\mathbf{k}'}^\dagger \hat{X}_{\mathbf{k}_1}^{\dagger (n_1^l-1)}    \hat{X}_{\mathbf{k}_2}^{\dagger n_2^l} .. \hat{X}_{\mathbf{k}_S}^{\dagger n_S^l} |\text{\o} \rangle \approx \\* \nonumber \approx &&\frac{2 n_1^l}{N_s} g_{\mathbf{k}'} \langle \hat{c}_{\mathbf{k}'} \rangle \sum_l \rho^X_{ll},
\end{eqnarray}
\end{widetext}
where in the last equality we considered the dominant contribution which appears for $\mathbf{k} = \mathbf{k}'$, ket-state reduces to the original $|N_l \rangle$, and we further note that $\sum_l \rho^X_{ll} = 1$ due to normalization. While other contributions may appear for $\mathbf{k} \neq \mathbf{k}'$, this shall happen in higher orders of $N_s^{-1}$. Finally, performing the same commutation for different momentum modes, we can write the diagonal contribution as
\begin{eqnarray}
&&\sum_{\mathbf{k}} g_{\mathbf{k}} \langle \hat{c}_{\mathbf{k}} \rangle \sum_{l} \rho^X_{ll} \langle N_{l}| \left[ \hat{X}_{\mathbf{k}'}, \hat{X}_{\mathbf{k}}^\dagger \right] | N_l \rangle = \\* \nonumber =&& g_{\mathbf{k}'} \langle \hat{c}_{\mathbf{k}'} \rangle - 2 g_{\mathbf{k}'} \langle \hat{c}_{\mathbf{k}'} \rangle \sum_l \rho^X_{ll} \frac{(n_1^l + n_2^l + .. + n_S^l)}{N_s} = \\* \nonumber =&& g_{\mathbf{k}'} \langle \hat{c}_{\mathbf{k}'} \rangle \Big( 1 - \frac{2 N^X_{\mathrm{tot}}}{N_s} \Big),
\end{eqnarray}
corresponding to the Eq.~\eqref{eq:Rabi_fin} in the main text.
The off-diagonal terms can be written in the same way. However, as they are proportional to coherences, this term becomes non-zero only after the threshold, and we generally can neglect it in the lowest order.

\end{document}